\documentclass[prl,twocolumn,superscriptaddress]{revtex4-1}
\usepackage{amsmath,amssymb,mathrsfs}
\usepackage{graphicx}
\usepackage{tikz}

\newcommand{\ua}{\uparrow}
\newcommand{\da}{\downarrow}

\def\be{\begin{equation}}       
\def\ee{\end{equation}}
\def\bea{\begin{eqnarray}}      
\def\eea{\end{eqnarray}}

\begin{document}

\title{Spin-orbital exchange of strongly interacting fermions on the 
$p$-band of a two-dimensional optical lattice}

\author{Zhenyu Zhou}
\affiliation{Department of Physics and Astronomy, University of
Pittsburgh, Pittsburgh, PA 15260}
\affiliation{
School of Physics, Astronomy and Computational Sciences, George Mason University, Fairfax, VA 22030}

\author{Erhai Zhao}
\affiliation{
School of Physics, Astronomy and Computational Sciences, George Mason University, Fairfax, VA 22030}

\author{W. Vincent Liu}
\affiliation{Department of Physics and Astronomy, University of
Pittsburgh, Pittsburgh, PA 15260}
\affiliation{
Wilczek Quantum Center, Zhejiang University of Technology, Hangzhou 310023, China
}
\begin{abstract}
Mott insulators with both spin and orbital degeneracy are pertinent to a large number of transition metal oxides. The intertwined spin and orbital fluctuations can lead to rather exotic phases such as quantum spin-orbital liquids. Here we consider two-component (spin 1/2) fermionic atoms with strong repulsive {interactions} on the $p$-band of the optical square lattice. We derive the spin-orbital exchange for quarter filling of the $p$-band {when the density fluctuations are suppressed}, and show it frustrates the development of long range spin order. Exact diagonalization indicates a spin-disordered ground state with ferro-orbital order. The system dynamically decouples into individual Heisenberg spin chains, each realizing a Luttinger liquid accessible at higher temperatures compared to atoms confined to the $s$-band.
\end{abstract}

\maketitle


{Quantum gases of ultracold atoms have served successfully as quantum simulators} of important superfluid and spin models derived from condensed matter.  A much less explored {potential} is to use {them} to gain deeper understanding of many-body orbital correlations. Electronic materials such as transitional metal oxides have shown intriguing phases where the role of orbital is found crucial~\cite{tokura_orbital_2000, hotta_orbital_2006}. In Mott insulators with degenerate $d$-orbitals, charge fluctuations are frozen by the strong Coulomb repulsion. At low energies, the spin and orbital degrees of freedom of neighboring sites are coupled by spin-orbital superexchange \cite{fazekas_lecture_1999}. A well known example  is the Kugel-Khomskii (KK) model for $e_g$ orbitals \cite{kugel_jahn-teller_1982}.
Often the spin-orbital exchange is frustrated, i.e., the exchange energy cannot be minimized simultaneously for all the bonds joining at the same site. Orbital degeneracy tends to enhance quantum fluctuations and suppress long-range order \cite{feiner_quantum_1997,PhysRevB.56.R14243}, thus providing an alternative route to realize exotic magnetic order or quantum spin liquids \cite{balents_spin_2010}. For example, there is strong theoretical evidence that the ground state of the $SU(4)$ symmetric KK model on the honeycomb lattice is a disordered quantum spin-orbital liquid \cite{corboz_spin-orbital_2012}. {From this perspective, it would be great to engineer a physical system to realize and probe such spin-orbital exchange models without the complication from other degrees of freedom such as lattice vibrations.}

Motivated by experiments on the higher orbital bands of optical lattices \cite{muller_state_2007, wirth_evidence_2011, soltan-panahi_quantum_2012,olschlager_unconventional_2011,olschlager_topologically_2012, olschlager_interaction-induced_2013}, we examine {the possibility of realizing} spin-orbital exchange for strongly interacting atoms on the $p$-band of two-dimensional (2D) optical lattice at commensurate fillings (i.e., the Mott limit). Due to the specific symmetries of the $p$-orbitals and the atomic interactions, {we expect that} the spin-orbital exchange of $p$-band fermions acquires a few unique features to distinct it from the KK exchange of $d$-orbital electrons with Coulomb interaction. Our main goal is to find the resultant spin and orbital long-range order, or the lack thereof, in simple {optical lattice} settings achievable in experiments. Previously, the orbital exchange for single component (spinless) fermions on the $p$-band has been discussed by two of us \cite{zhao_orbital_2008} and Wu \cite{wu_orbital_2008}. The work on two-component (spin 1/2) $p$-band fermions has largely focused on spin-only models and the ferromagnetic or antiferromagnetic long-range order, for example, for the the half filled cubic lattice \cite{wu_theory_2008} and various fillings of 2D lattices \cite{wang_magnetism_2008, zhang_proposed_2010, li_exact_2014}. 

In this letter we focus on $1/4$ filling of the $p$-band, {where density fluctuations are suppressed by repulsive interactions between fermions with either the same or opposite spin, and derive the effective exchange interaction between the orbital and spin degrees of freedom}.  We show that locally for an individual bond, the spin-orbital exchange prefers the alignment of the $p$-orbitals and the formation of spin singlet. Such lowest energy configuration apparently cannot be achieved for all the bonds at once. To partially alleviate the frustration, the system settles into a spin-disordered ground state with ferro-orbital order that is  spatially organized into chains. This conjecture is supported by exact diagonalization of finite systems with various sizes and boundary conditions. Such quasi-one-dimensional spin liquid is in dramatic contrast to the long-range magnetic order of $p$-band fermions predicted for other regimes such as half filling \cite{wu_theory_2008}. Our results {indicate} that $p$-band fermions, and more generally spin-orbital exchange of ultracold atoms, offer rich possibilities for novel states of matter. 

First we show how the spin-orbit exchange { can arise} from the microscopic Hamiltonian of interacting atoms on optical lattice. For simplicity, consider a 2D optical lattice on the $xy$ plane, with a lattice depth $V$ much larger than the recoil energy $E_R$. The lattice potential at each lattice site is then well approximated by a 2D harmonic oscillator of frequency $\omega$. The Wannier functions are approximated by the corresponding wave functions of the harmonic oscillator: the ground state $s$ orbital, the doubly degenerate first excited state $p_x$ and $p_y$ (or $x$ and $y$ for short) orbital, etc. The excitation energy from the $s$ to the $p$ orbital is $\hbar\omega$.
The $s$-wave scattering between two hyperfine species of fermionic atoms, referred to as spin up and down, is well described by a contact interaction. We assume it is repulsive and its strength is controlled by tuning magnetic field around a Feshbach resonance.
Expanding the fermion field operator in the Wannier basis and computing the direct 
and exchange integrals using the wave functions of the $s$ and $p$ orbitals, the interaction Hamiltonian for each site becomes
\begin{align*}
H_{A}&=Un_{s\ua}n_{s\da}+\frac{3U}{4}[n_{x\ua}n_{x\da}+n_{y\ua}n_{y\da}] +\frac{U}{4}[n_{x\ua}n_{y\da} \\ 
&+n_{y\ua}n_{x\da} +\Delta^\dagger_{x}\Delta_y+\Delta^\dagger_{y}\Delta_x -S^+_xS^-_y-S_y^+S^-_x] + ...
\end{align*}
where the ellipsis includes terms coupling the $s$ and $p$ orbitals, and terms involving higher orbitals.
Here, $n_{\mu,\sigma}=c^\dagger_{\mu,\sigma}c_{\mu,\sigma}$, $S^+_\mu=c^\dagger_{\mu,\ua}c_{\mu,\da}$, $\Delta_\mu=c_{\mu,\ua}c_{\mu,\da}$, and $c^\dagger_{\mu\sigma}$ is the fermion creation operator for orbital $\mu=s,x,y$ and spin $\sigma=\ua,\da$.  
The onsite interaction energy is $U>0$ for two atoms in the $s$ orbital, and $3U/4$ for two atoms in the $p_x$ and the $p_y$ orbital respectively. We observe that besides the density interactions ($n_{\ua}n_{\da}$), Hund's rule coupling ($S^+ S^-$) and pair transfer ($\Delta^\dagger\Delta$) terms are of the same order and equally important \cite{fazekas_lecture_1999}. 

{We assume that there is a large onsite repulsive interaction $U'$ between fermions of the same spin, $U'\gg U$. It forbids two fermions of the same spin from occupying the same site (e.g., one occupying the $p_x$ orbital and the other occupying $p_y$). If $U'$ is absent or weak, fermions can hop around resulting a metallic state with ferromagnetic long range order \cite{li_exact_2014}, instead of a Mott state. The ferromagnetic ground state has been proved rigorously in the limit of $U\rightarrow \infty$ in Ref. \cite{li_exact_2014} and is conjectured to hold also for finite $U$ \cite{priv}. It is challenging, but in principle feasible, to achieve a large $U'$ experimentally. One way is to use an optical Feshbach resonance \cite{PhysRevLett.77.2913,PhysRevA.56.1486,PhysRevLett.85.4462,PhysRevLett.93.123001,PhysRevLett.105.050405,PhysRevLett.107.073202} to tune the $p$-wave interaction, as theoretically proposed in Ref. \cite{PhysRevLett.103.023201,PhysRevA.82.062704} and experimentally demonstrated in Ref. \cite{PhysRevA.87.010704}. Large $p$-wave interaction was also assumed for spinless $p$-orbital fermions in previous work \cite{zhao_orbital_2008, wu_orbital_2008, PhysRevE.84.061127, PhysRevA.84.051603}. 

We focus on the case of three atoms per site. Without interaction ($U$=0), two atoms of opposite spin fill the $s$ orbital, and the third atom occupies either the $p_x$ or $p_y$ orbital, corresponding to quarter filling of the $p$-band. In the presence of onsite interaction $H_A$, diagonalization of $H_A$ shows that as long as $U<U_c=\hbar \omega/1.4$, the ground state configuration remains roughly the same. The probability for the $p_x$ (or $p_y$) orbital to be doubly occupied due to interaction is less than $4.2\%$. In what follows, we shall assume $U\ll \hbar\omega$. Then the doubly occupied $s$ orbital is well separated from the $p$ orbital in energy. It will remain dynamically inert and can be safely neglected. Then $H_A$ reduces to a $p$-orbital only Hamiltonian taking the following compact form, 
\[
H_a=-\frac{U}{8}[L^2_z+4\vec{S}^2]+\frac{3U}{8}(n_x+n_y).
\]
Here the spin and angular momentum operator are defined as $\vec{S}=\frac{1}{2}c^{\dagger}_{\mu,\sigma}\boldsymbol{\sigma}_{\sigma,\sigma'}c_{\mu,\sigma'}$, $
L_z=(-i)[c^\dagger_{x,\sigma}c_{y,\sigma}-h.c.]$, and $n_x=n_{x,\ua}+n_{y,\da}$ \cite{liu_atomic_2006}. Repeated indices, $\mu=x,y$ and $\sigma=\ua,\da$, are summed over, and $\boldsymbol{\sigma}$ is the Pauli matrix. With only one fermion on the $p$-orbital, the ground state is four-fold degenerate. We introduce a graphic notation for these four states (see Fig. 1). The upper (lower) semi circle denotes the $p_x$ ($p_y$) orbital, and an up (down) arrow indicates the orbital is occupied by an atom with spin up (down).

Besides the onsite interaction $H_a$ proportional to $U$, the $p$-orbital fermions can also hop. On the square lattice, the leading process is the longitudinal hopping,
\[
H_t=-t\sum_{i,\sigma}c^\dagger_{x,\sigma}(i)c_{x,\sigma}(i+\hat{x})+c^\dagger_{y,\sigma}(i)c_{y,\sigma}(i+\hat{y})+h.c.
\]
Namely, $p_x$ ($p_y$) fermions only hop along the $x$ ($y$) axis between nearest neighbors. Here $i$ labels the site and is the short-hand notation for the lattice vector $\mathbf{R}_i$ with the lattice spacing set to 1. We neglect transverse hopping, e.g., $p_x$ fermions hopping in the $y$ direction. 
Its magnitude is only $t/8$ for $V=5E_R$ and further decreases as $V/E_R$ increases. 

The total Hamiltonian for the $p$-orbital fermions then is $\sum_i H_a(i)+H_t$. We focus on Mott states corresponding to quarter filling of the $p$ band with equal spin populations in the strongly correlated regime {$U'\gg U\gg t$}. The large onsite repulsion suppresses density fluctuations. In the lowest order approximation, $H_t$ can be neglected so the system decouples into individual sites, each described by $H_a$. Its ground state has a massive degeneracy $4^N$ where $N$ is the number of sites. $H_t$ appears as a perturbation to the atomic Hamiltonian $\sum_i H_a(i)$. Virtual hopping processes give rise to spin-orbital exchange interaction between neighboring sites. The spin-orbital exchange can be obtained by standard second order perturbation theory \cite{oles_quantum_2000}. It lifts the degeneracy and dictates the spin and/or orbital order within the Mott state. 

\begin{figure}
\includegraphics[width=3.2in]{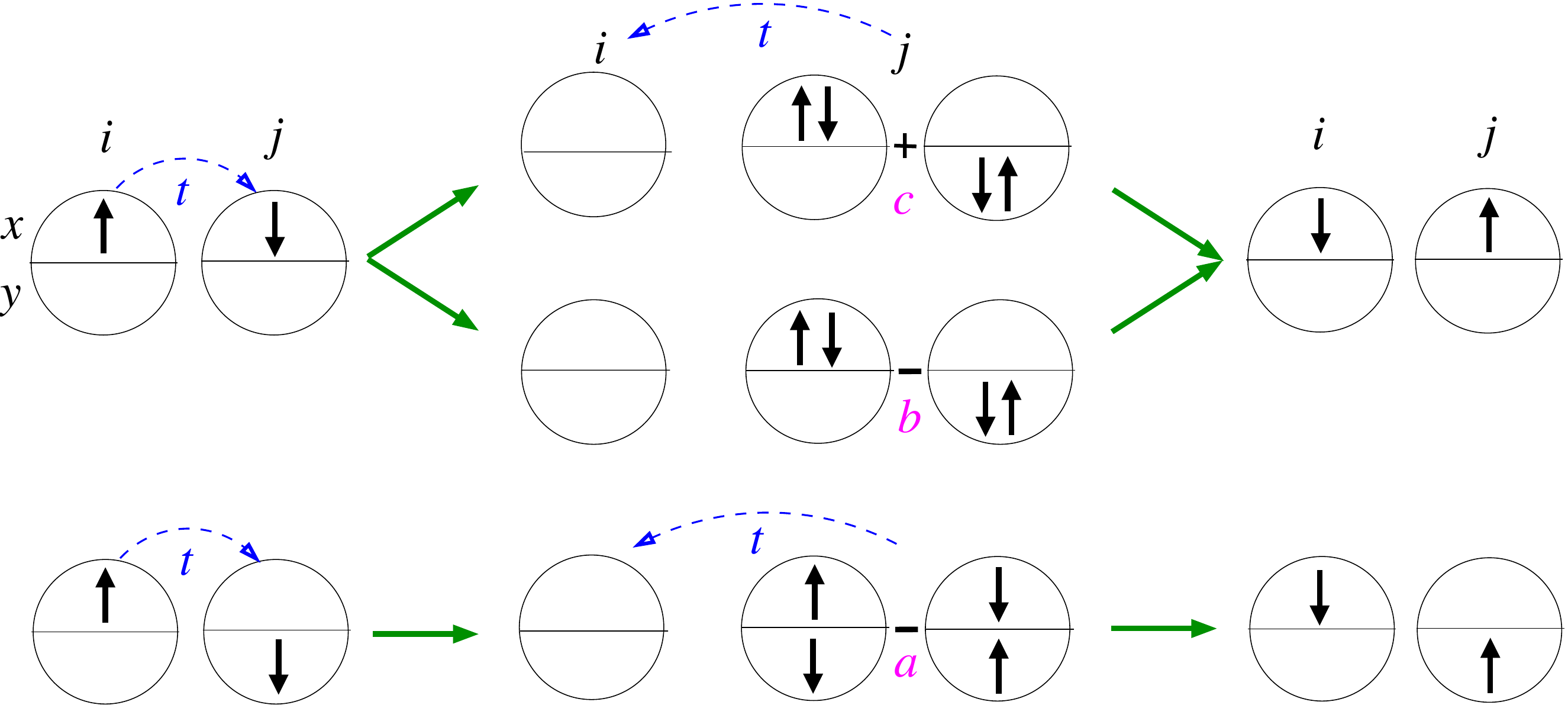}
\caption{Virtual hopping processes giving rise to the spin-orbital exchange. $i$ and $j$ label two neighboring sites. An arrow in the upper/lower semicircle means the $p_x$/$p_y$ orbital is occupied by atoms of given spin. $a, b, c$ are intermediate states for two atoms on the same site $j$. }
\label{f1}
\end{figure} 

First consider a bond along the $x$ direction connecting site $i$ and $j=i+\hat{x}$. As shown schematically in Fig. \ref{f1}, in the initial state, each site has one fermion in the $p$-orbital. Hopping of a $p_x$ fermion, say from $i$ to $j$, creates an intermediate state $|n\rangle$ with two fermions on site $j$. Diagonalization of $H_a$ shows that there are three such eigenstates, $n=a, b, c$ (see Fig. \ref{f1}), with excitation energy $\epsilon_{a,b}=U/2$, and $\epsilon_c=U$.  Note that the $p_x$ fermion has to hop back to its initial position site $i$ from the intermediate state, because $p_y$ fermion cannot hop in the $x$ direction. In addition, the {exchange interaction} is restricted to the singlet channel {(the exchange in the triplet channel is on the order of $t^2/U'$, which is negligible)}. Thus, the spin-orbital exchange is most easily obtained by using projection operators,
\be
H^i_x=-\sum_{n=a,b,c}\sum_{\mu,\nu}\frac{t^2}{\epsilon_n}(\frac{1}{4}-\vec{S}_i\cdot \vec{S}_j)P_{i\mu}P_{j\nu}.
\ee
Here $\mu,\nu=x,y$ denotes the initial orbital state of site $i$ and $j$ respectively, and $\epsilon_n$ is excitation energy of the intermediate state $|n\rangle$. $P_{i\mu}$ is the orbital projection operator corresponding to state $|i\mu\rangle$, i.e., one fermion in orbital $\mu$ at site $i$, 
\begin{align}
P_{ix}&\equiv|ix\rangle\langle ix|\equiv 1/2+\tau^z_i, \nonumber \\
P_{iy}&\equiv|iy\rangle\langle iy|\equiv 1/2-\tau^z_i,
\end{align} 
where we also introduced the pseudospin operator $\tau_z$ in the orbital space. $({1}/{4}-\vec{S}_i\vec{S}_j)$ is the projector operator onto the spin singlet channel. Collecting terms, we obtain
\be
H^i_x=\frac{t^2}{U}(\vec{S}_i\cdot \vec{S}_j-\frac{1}{4})( \frac{5}{2}+3\tau^z_i+3\tau^z_j+2\tau^z_i\tau^z_j).
\ee
This is one of the central results of this paper. By symmetry, the exchange along bonds in the $y$ direction, $j=i+\hat{y}$, 
\be
H^i_y=\frac{t^2}{U}(\vec{S}_i\cdot \vec{S}_j-\frac{1}{4})( \frac{5}{2}-3\tau^z_i-3\tau^z_j+2\tau^z_i\tau^z_j).
\ee
The low energy effective Hamiltonian for the $p$-band fermions is the sum of the spin-orbit exchange for all the bonds on the square lattice,
\be
H_{so}=\sum_{i} \left[H_x^i +H_y^i\right]. 
\ee

It is illuminating to compare $H_{so}$ with the $SU(4)$ symmetric KK model \cite{li_su4_1998, yamashita_su4_1998, wang_z2_2009, corboz_spin-orbital_2012}, where the exchange takes the form $(\vec{S}_i\cdot\vec{S}_j+1/4)(\vec{\tau}_i\cdot\vec{\tau}_j+1/4)$, or the original KK model \cite{kugel_crystal_1973, kugel_jahn-teller_1982} for $e_g$ electrons where the exchange along the three cubic axes ($a,b,c$) involves different pseudo-spin operators, $\tau^{a/b}=(\pm\sqrt{3}\tau^x-\tau^z)/2$ and $\tau^c=\tau^z$. Here, only $\tau_z$ appears in $H_{so}$. The coupling is Ising-like in the orbital sector but Heisenberg-like in the spin sector. $H_{so}$ has discrete symmetry $\tau_z\rightarrow -\tau_z$ corresponding to C$_4$ rotation, $x\rightarrow y$.  Gorshkov et al. \cite{gorshkov_two-orbital_2010} have proposed that KK-type models can be engineered using alkaline earth atoms, where two electronic states of atoms play the role of orbitals. Here in $H_{so}$ the orbital refers to the Wannier orbital of atoms on lattice, as in the original KK model, rather than its internal electronic states.

\begin{figure}
\includegraphics[width=3.4in]{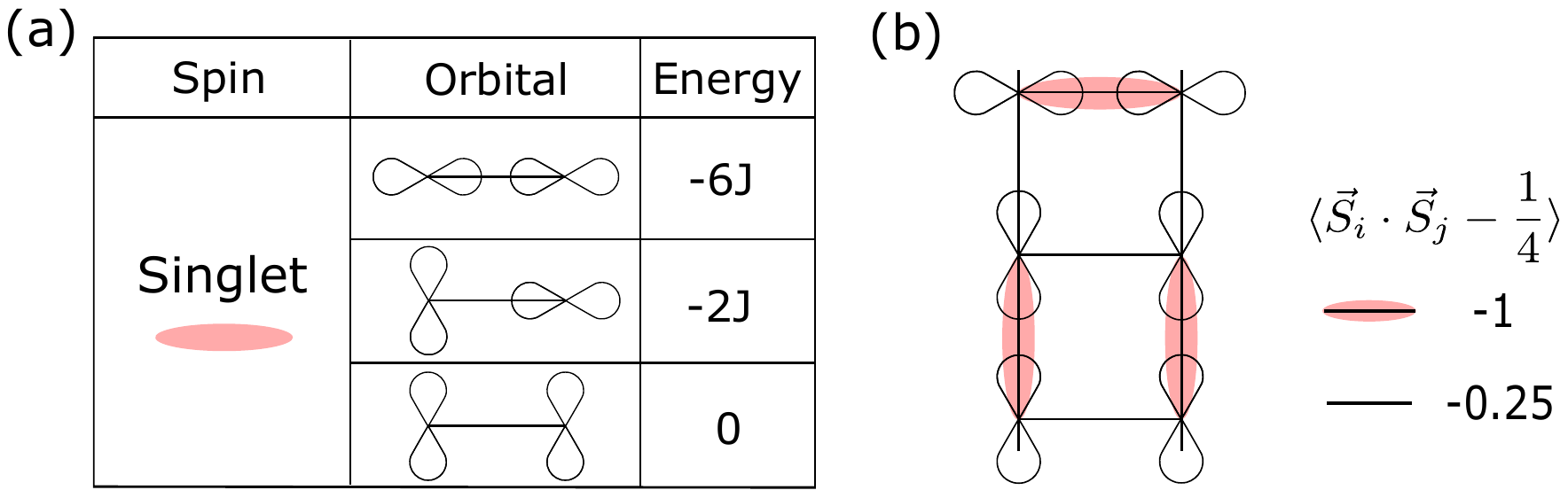}
\caption{(a) The eigenstates of $H^i_x$ for a single bond. (b) One of the degenerate ground state of a two-leg ladder. The value of the nearest neighbor spin correlation is shown graphically.}
\label{f2}
\end{figure} 

In the remainder of this paper, we focus on the ground state and the low energy excitations of $H_{so}$. 
We first consider a single horizontal bond described by $H^i_x$. Its ground state is a spin singlet and orbital triplet with both orbitals  aligned in the $x$ direction, $|\psi_{x}\rangle= \frac{1}{\sqrt{2}} (|i\ua\rangle |j\downarrow\rangle 
- |i\da\rangle |j\ua\rangle)\otimes |ix\rangle|jx\rangle$. As shown in  Fig. 2(a), the ground state energy is $E_d=-6J$, where $J\equiv t^2/U$ is the energy unit. Other orbital configurations within the spin singlet sector have much higher energy. 
The ground state for a vertical bond along $y$, $|\psi_{y}\rangle$, is obtained from $|\psi_{x}\rangle$ by the replacement $x\rightarrow y$. We shall refer to local states $|\psi_{x/y}\rangle$ as dimers and represent them graphically as shaded ovals. They have characteristic spin correlation  $\langle\vec{S}_i\cdot\vec{S}_j-1/4\rangle=-1$. Clearly, on the square lattice, the $x$ and $y$ bonds joining at a lattice site cannot minimize their energies to $E_d$ simultaneously. This is a classic syndrome of frustration, which is quite common in spin-orbital exchange models. Out of the four bonds connected to the same site, only one can form a dimer. 
Take a $2\times 2$ cluster (a plaquette)  with open boundary condition for example. Exact diagonalization (ED) shows that the ground state is two-fold degenerate with energy $-12J$.  It has ferro-orbital order $\prod_i |ix\rangle$ (or $\prod_i |iy\rangle$, see the bottom plaquette of Fig. 2(b)). Two dimers, i.e. spin singlets, are formed on the two bonds parallel to the aligned orbitals, each achieving its lowest energy $E_d$, leaving the remaining two bonds frustrated. Similarly, Fig. 2(b) shows one of the degenerate ground states of a 2$\times$3 cluster with periodic boundary condition in the $y$ direction and open boundary condition in the $x$ direction. The orbital and spin configuration also correspond to a dimer covering of the lattice. However, our systematic ED analysis of $H_{so}$ for bigger clusters rejects dimer covering, and picks a state with ferro-orbital long range order, as the ground state of $H_{so}$ for the infinite lattice. 

\begin{figure}
\includegraphics[width=3.2in]{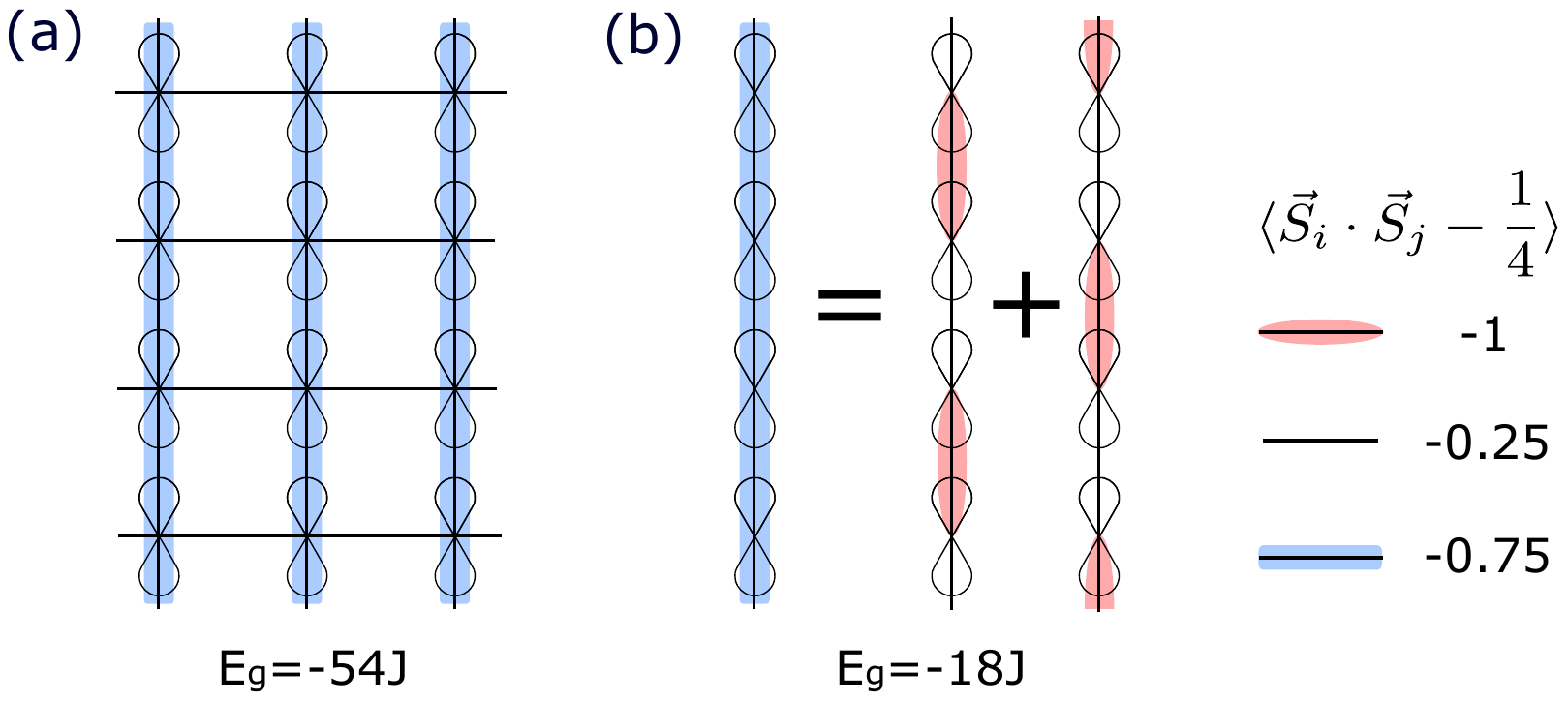}
\caption{ (a) The ground state of $H_{so}$ for a $3\times 4$ cluster with periodic boundary conditions. (b) The ground state of a single chain ($1\times 4$) with periodic boundary condition in the $y$ direction. The ground state energy $E_g$ is measured in $J=t^2/U$.}
\label{f3}
\end{figure} 

For instance, Fig. 3(a) shows the unique ground state of a $3\times 4$ cluster with periodic boundary conditions. It has ferro-orbital order with $\langle \tau^z_{i} \rangle=1/2$ for all the sites, i.e., all orbitals aligning along $y$. There is however no spin order, $\langle S^z_{i} \rangle=0$. The spin correlation $\langle\vec{S}_i\cdot\vec{S}_j-1/4\rangle$ takes the value of $-1/4$ for all horizontal bonds (thin lines) and $-3/4$ for vertical bonds (thick blue lines). Such correlation indicates that the cluster decouples into three vertical chains. Fig. 3(b) shows the ground state of an individual chain containing 4 sites with periodic boundary condition in the $y$ direction. According to ED, it also has ferro-orbital order, and the ground state wave function is the equal amplitude superposition of two dimer coverings as graphically depicted in Fig. 3(b). The ground state energy of the $3\times 4$ cluster is exactly three times of the single chain. We have verified that its ground state wave function is nothing but the direct product of those of the three individual chains.  In comparison, a dimer covering as a trial state can only yield an energy expectation value as low as $-3.75J$ per site, much higher than $-4.5J$ of the ED ground state above. Similarly, a mean field variational calculation of $H_{so}$ assuming N\'eel order of spins predicts ferro-orbital order but yields an even higher energy of $-3J$ per site. 
The development of ferro-orbital order and the decoupling of the cluster into one-dimensional (1D) chains are also observed for two-leg ($2\times 4$ and $2\times 6$) and three-leg ($3\times 4$) ladders with $y$-periodic boundary conditions. Fig. 4 summarizes the ground state energy per site $E_g/N$ for 1D chains, two-leg ladders, and the $3\times 4$ cluster. The value of $E_g/N$ is identical, e.g., for the $3\times 4$, $2\times 4$, and $1\times 4$ cluster, revealing the decoupling of the ladder/cluster into chains.

\begin{figure}
\includegraphics[width=3in]{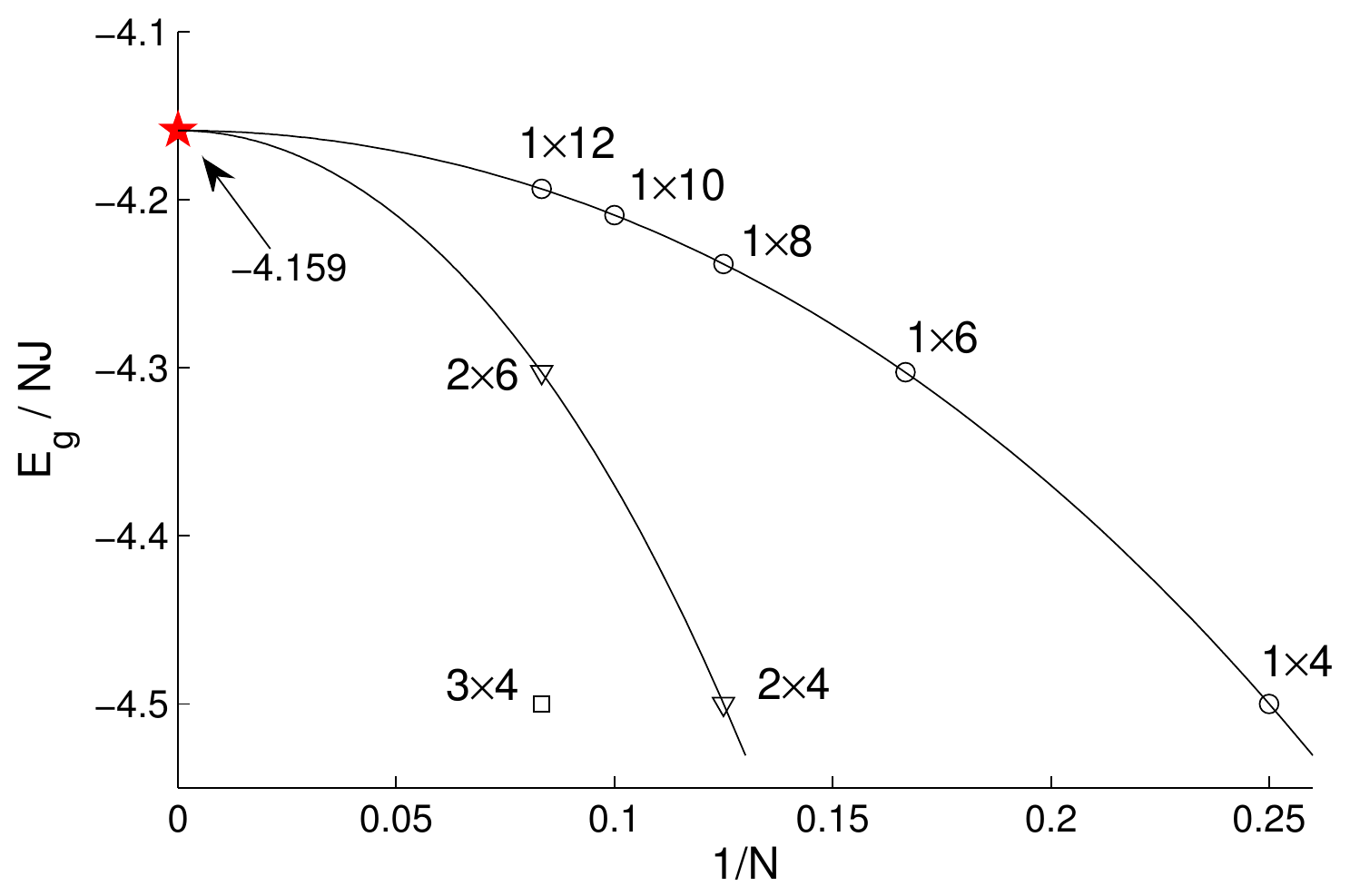}
\caption{The ground state energy per site, $E_g/N$,  in unit of $J$ obtained by exact diagonalization of $H_{so}$ for different clusters. Finite size scaling yields $E_g/N=-4.159J$ (filled star) in the thermodynamic limit $N\rightarrow \infty$.}
\label{f4}
\end{figure}

From the evidences above, we infer that spin-orbital exchange favors ferro-orbital order on the square lattice, where the $p$-orbitals at all sites align in the $x$ (or $y$) direction. At low temperatures, $T<J$, the 2D system dynamically decouples into 1D chains. With the orbital degree of freedom frozen out, each chain is described by a spin $1/2$ anti-ferromagnetic Heisenberg Hamiltonian
\be
H_{1D}= 6J \sum_i (\vec{S}_i\cdot \vec{S}_{i+1}-\frac{1}{4}).
\ee
This 1D model is exactly solvable by Bethe Ansatz \cite{bethe_zur_1931}. 
Finite size scaling of our ED results by fitting $E_g/N$ to polynomials of $1/N$ indeed shows $E_g/N$ extrapolate to $-(\ln 2)6J=-4.159J$ as $N\rightarrow \infty$, in excellent agreement with Bethe ansatz (see Fig. 4).  As well known, there is no long-range spin order for the 1D Heisenberg model, and its low energy effective model is a Luttinger liquid featuring algebraically decaying spin correlation functions. The orbital excitations are gapped, but the spin excitations are gapless and highly anisotropic. The elementary excitations are spinons traversing in the direction of the ordered orbitals. Higher order effects such as the small transverse hopping of $p$-orbitals neglected so far will introduce coupling between chains, and potentially long-range spin order at extremely low temperatures, $T\ll J$. For a broad temperature window below $J$, experiments will access the properties of Luttinger liquids.
Compared to 1D Hubbard (or Heisenberg) model based on $s$-band fermions, the hopping of $p$-band fermions, and accordingly the exchange scale $J$, is significantly enhanced. This is beneficial for the experimental exploration of the physics beyond the Luttinger liquid paradigm \cite{RevModPhys.84.1253}, the quantum dynamics \cite{kinoshita2006quantum,rigol2008thermalization} and dimensional crossover of 1D antiferromagnets \cite{PhysRevLett.77.4446,0305-4470-29-11-003}.

The spin-disordered ground state of $H_{so}$ found here is not as exotic as quantum spin liquids \cite{balents_spin_2010} in 2D with topological order or fractional statistics. Despite this, it serves as a dramatic, unprecedented example of how orbital order enhances quantum fluctuations to prevent spin order and lead to dimension reduction in a quantum gas. It is similar in spirit to the transition metal oxide Tl$_2$Ru$_2$O$_7$ conjectured to self-organize into zigzag spin 1 chains \cite{lee_spin_2006, van_den_brink_transition_2006} 
due to orbital order at low temperatures. We stress that spin-orbital exchange of $p$-band fermions acquires new features that are closely tied to the $p$-orbital symmetry and the specific forms of interaction for cold atoms. Our work represents the first step to understand this new form of spin-orbital exchange. $H_{x/y}$ can be generalized to find  $H_{so}$ for other 2D lattices, such as the triangular and hexagonal lattices, by orbital rotations \cite{zhao_orbital_2008}. Finding their ground state orbital and spin order is a challenging open problem for future work.   

{We thank Congjun Wu and Haiyuan Zou for helpful discussions}.
This work is supported by AFOSR FA9550-12-1-0079 (ZZ, EZ, and WVL), NSF PHY-1205504 (EZ),  and jointly by ARO W911NF-11-1-0230, DARPA OLE Program through ARO, The Pittsburgh Foundation and its Charles E. Kaufman Foundation, and Overseas Scholar Collaboration Program of NSF of China No. 11429402 sponsored by Peking University (WVL). 
\bibliography{spin_orbital}

\begin{thebibliography}{47}%
\makeatletter
\providecommand \@ifxundefined [1]{%
 \@ifx{#1\undefined}
}%
\providecommand \@ifnum [1]{%
 \ifnum #1\expandafter \@firstoftwo
 \else \expandafter \@secondoftwo
 \fi
}%
\providecommand \@ifx [1]{%
 \ifx #1\expandafter \@firstoftwo
 \else \expandafter \@secondoftwo
 \fi
}%
\providecommand \natexlab [1]{#1}%
\providecommand \enquote  [1]{``#1''}%
\providecommand \bibnamefont  [1]{#1}%
\providecommand \bibfnamefont [1]{#1}%
\providecommand \citenamefont [1]{#1}%
\providecommand \href@noop [0]{\@secondoftwo}%
\providecommand \href [0]{\begingroup \@sanitize@url \@href}%
\providecommand \@href[1]{\@@startlink{#1}\@@href}%
\providecommand \@@href[1]{\endgroup#1\@@endlink}%
\providecommand \@sanitize@url [0]{\catcode `\\12\catcode `\$12\catcode
  `\&12\catcode `\#12\catcode `\^12\catcode `\_12\catcode `\%12\relax}%
\providecommand \@@startlink[1]{}%
\providecommand \@@endlink[0]{}%
\providecommand \url  [0]{\begingroup\@sanitize@url \@url }%
\providecommand \@url [1]{\endgroup\@href {#1}{\urlprefix }}%
\providecommand \urlprefix  [0]{URL }%
\providecommand \Eprint [0]{\href }%
\providecommand \doibase [0]{http://dx.doi.org/}%
\providecommand \selectlanguage [0]{\@gobble}%
\providecommand \bibinfo  [0]{\@secondoftwo}%
\providecommand \bibfield  [0]{\@secondoftwo}%
\providecommand \translation [1]{[#1]}%
\providecommand \BibitemOpen [0]{}%
\providecommand \bibitemStop [0]{}%
\providecommand \bibitemNoStop [0]{.\EOS\space}%
\providecommand \EOS [0]{\spacefactor3000\relax}%
\providecommand \BibitemShut  [1]{\csname bibitem#1\endcsname}%
\let\auto@bib@innerbib\@empty
\bibitem [{\citenamefont {Tokura}\ and\ \citenamefont
  {Nagaosa}(2000)}]{tokura_orbital_2000}%
  \BibitemOpen
  \bibfield  {author} {\bibinfo {author} {\bibfnamefont {Y.}~\bibnamefont
  {Tokura}}\ and\ \bibinfo {author} {\bibfnamefont {N.}~\bibnamefont
  {Nagaosa}},\ }\href {\doibase 10.1126/science.288.5465.462} {\bibfield
  {journal} {\bibinfo  {journal} {Science}\ }\textbf {\bibinfo {volume}
  {288}},\ \bibinfo {pages} {462} (\bibinfo {year} {2000})}\BibitemShut
  {NoStop}%
\bibitem [{\citenamefont {Hotta}(2006)}]{hotta_orbital_2006}%
  \BibitemOpen
  \bibfield  {author} {\bibinfo {author} {\bibfnamefont {T.}~\bibnamefont
  {Hotta}},\ }\href {\doibase 10.1088/0034-4885/69/7/R02} {\bibfield  {journal}
  {\bibinfo  {journal} {Reports on Progress in Physics}\ }\textbf {\bibinfo
  {volume} {69}},\ \bibinfo {pages} {2061} (\bibinfo {year}
  {2006})}\BibitemShut {NoStop}%
\bibitem [{\citenamefont {Fazekas}(1999)}]{fazekas_lecture_1999}%
  \BibitemOpen
  \bibfield  {author} {\bibinfo {author} {\bibfnamefont {P.}~\bibnamefont
  {Fazekas}},\ }\href@noop {} {\emph {\bibinfo {title} {Lecture Notes on
  Electron Correlation and Magnetism}}}\ (\bibinfo  {publisher} {World
  Scientific},\ \bibinfo {year} {1999})\BibitemShut {NoStop}%
\bibitem [{\citenamefont {Kugel'}\ and\ \citenamefont
  {Khomski\u{i}}(1982)}]{kugel_jahn-teller_1982}%
  \BibitemOpen
  \bibfield  {author} {\bibinfo {author} {\bibfnamefont {K.~I.}\ \bibnamefont
  {Kugel'}}\ and\ \bibinfo {author} {\bibfnamefont {D.~I.}\ \bibnamefont
  {Khomski\u{i}}},\ }\href {\doibase 10.1070/PU1982v025n04ABEH004537}
  {\bibfield  {journal} {\bibinfo  {journal} {Soviet Physics Uspekhi}\ }\textbf
  {\bibinfo {volume} {25}},\ \bibinfo {pages} {231} (\bibinfo {year}
  {1982})}\BibitemShut {NoStop}%
\bibitem [{\citenamefont {Feiner}\ \emph {et~al.}(1997)\citenamefont {Feiner},
  \citenamefont {Ole\'{s}},\ and\ \citenamefont
  {Zaanen}}]{feiner_quantum_1997}%
  \BibitemOpen
  \bibfield  {author} {\bibinfo {author} {\bibfnamefont {L.~F.}\ \bibnamefont
  {Feiner}}, \bibinfo {author} {\bibfnamefont {A.~M.}\ \bibnamefont
  {Ole\'{s}}}, \ and\ \bibinfo {author} {\bibfnamefont {J.}~\bibnamefont
  {Zaanen}},\ }\href {\doibase 10.1103/PhysRevLett.78.2799} {\bibfield
  {journal} {\bibinfo  {journal} {Physical Review Letters}\ }\textbf {\bibinfo
  {volume} {78}},\ \bibinfo {pages} {2799} (\bibinfo {year}
  {1997})}\BibitemShut {NoStop}%
\bibitem [{\citenamefont {Khaliullin}\ and\ \citenamefont
  {Oudovenko}(1997)}]{PhysRevB.56.R14243}%
  \BibitemOpen
  \bibfield  {author} {\bibinfo {author} {\bibfnamefont {G.}~\bibnamefont
  {Khaliullin}}\ and\ \bibinfo {author} {\bibfnamefont {V.}~\bibnamefont
  {Oudovenko}},\ }\href {\doibase 10.1103/PhysRevB.56.R14243} {\bibfield
  {journal} {\bibinfo  {journal} {Phys. Rev. B}\ }\textbf {\bibinfo {volume}
  {56}},\ \bibinfo {pages} {R14243} (\bibinfo {year} {1997})}\BibitemShut
  {NoStop}%
\bibitem [{\citenamefont {Balents}(2010)}]{balents_spin_2010}%
  \BibitemOpen
  \bibfield  {author} {\bibinfo {author} {\bibfnamefont {L.}~\bibnamefont
  {Balents}},\ }\href {\doibase 10.1038/nature08917} {\bibfield  {journal}
  {\bibinfo  {journal} {Nature}\ }\textbf {\bibinfo {volume} {464}},\ \bibinfo
  {pages} {199} (\bibinfo {year} {2010})}\BibitemShut {NoStop}%
\bibitem [{\citenamefont {Corboz}\ \emph {et~al.}(2012)\citenamefont {Corboz},
  \citenamefont {Lajk\'{o}}, \citenamefont {L\"{a}uchli}, \citenamefont
  {Penc},\ and\ \citenamefont {Mila}}]{corboz_spin-orbital_2012}%
  \BibitemOpen
  \bibfield  {author} {\bibinfo {author} {\bibfnamefont {P.}~\bibnamefont
  {Corboz}}, \bibinfo {author} {\bibfnamefont {M.}~\bibnamefont {Lajk\'{o}}},
  \bibinfo {author} {\bibfnamefont {A.~M.}\ \bibnamefont {L\"{a}uchli}},
  \bibinfo {author} {\bibfnamefont {K.}~\bibnamefont {Penc}}, \ and\ \bibinfo
  {author} {\bibfnamefont {F.}~\bibnamefont {Mila}},\ }\href {\doibase
  10.1103/PhysRevX.2.041013} {\bibfield  {journal} {\bibinfo  {journal}
  {Physical Review X}\ }\textbf {\bibinfo {volume} {2}},\ \bibinfo {pages}
  {041013} (\bibinfo {year} {2012})}\BibitemShut {NoStop}%
\bibitem [{\citenamefont {M\"{u}ller}\ \emph {et~al.}(2007)\citenamefont
  {M\"{u}ller}, \citenamefont {F\"{o}lling}, \citenamefont {Widera},\ and\
  \citenamefont {Bloch}}]{muller_state_2007}%
  \BibitemOpen
  \bibfield  {author} {\bibinfo {author} {\bibfnamefont {T.}~\bibnamefont
  {M\"{u}ller}}, \bibinfo {author} {\bibfnamefont {S.}~\bibnamefont
  {F\"{o}lling}}, \bibinfo {author} {\bibfnamefont {A.}~\bibnamefont {Widera}},
  \ and\ \bibinfo {author} {\bibfnamefont {I.}~\bibnamefont {Bloch}},\ }\href
  {\doibase 10.1103/PhysRevLett.99.200405} {\bibfield  {journal} {\bibinfo
  {journal} {Physical Review Letters}\ }\textbf {\bibinfo {volume} {99}},\
  \bibinfo {pages} {200405} (\bibinfo {year} {2007})}\BibitemShut {NoStop}%
\bibitem [{\citenamefont {Wirth}\ \emph {et~al.}(2011)\citenamefont {Wirth},
  \citenamefont {\"{O}lschl\"{a}ger},\ and\ \citenamefont
  {Hemmerich}}]{wirth_evidence_2011}%
  \BibitemOpen
  \bibfield  {author} {\bibinfo {author} {\bibfnamefont {G.}~\bibnamefont
  {Wirth}}, \bibinfo {author} {\bibfnamefont {M.}~\bibnamefont
  {\"{O}lschl\"{a}ger}}, \ and\ \bibinfo {author} {\bibfnamefont
  {A.}~\bibnamefont {Hemmerich}},\ }\href {\doibase 10.1038/nphys1857}
  {\bibfield  {journal} {\bibinfo  {journal} {Nature Physics}\ }\textbf
  {\bibinfo {volume} {7}},\ \bibinfo {pages} {147} (\bibinfo {year}
  {2011})}\BibitemShut {NoStop}%
\bibitem [{\citenamefont {Soltan-Panahi}\ \emph {et~al.}(2012)\citenamefont
  {Soltan-Panahi}, \citenamefont {L\"{u}hmann}, \citenamefont {Struck},
  \citenamefont {Windpassinger},\ and\ \citenamefont
  {Sengstock}}]{soltan-panahi_quantum_2012}%
  \BibitemOpen
  \bibfield  {author} {\bibinfo {author} {\bibfnamefont {P.}~\bibnamefont
  {Soltan-Panahi}}, \bibinfo {author} {\bibfnamefont {D.-S.}\ \bibnamefont
  {L\"{u}hmann}}, \bibinfo {author} {\bibfnamefont {J.}~\bibnamefont {Struck}},
  \bibinfo {author} {\bibfnamefont {P.}~\bibnamefont {Windpassinger}}, \ and\
  \bibinfo {author} {\bibfnamefont {K.}~\bibnamefont {Sengstock}},\ }\href
  {\doibase 10.1038/nphys2128} {\bibfield  {journal} {\bibinfo  {journal}
  {Nature Physics}\ }\textbf {\bibinfo {volume} {8}},\ \bibinfo {pages} {71}
  (\bibinfo {year} {2012})}\BibitemShut {NoStop}%
\bibitem [{\citenamefont {\"{O}lschl\"{a}ger}\ \emph
  {et~al.}(2011)\citenamefont {\"{O}lschl\"{a}ger}, \citenamefont {Wirth},\
  and\ \citenamefont {Hemmerich}}]{olschlager_unconventional_2011}%
  \BibitemOpen
  \bibfield  {author} {\bibinfo {author} {\bibfnamefont {M.}~\bibnamefont
  {\"{O}lschl\"{a}ger}}, \bibinfo {author} {\bibfnamefont {G.}~\bibnamefont
  {Wirth}}, \ and\ \bibinfo {author} {\bibfnamefont {A.}~\bibnamefont
  {Hemmerich}},\ }\href {\doibase 10.1103/PhysRevLett.106.015302} {\bibfield
  {journal} {\bibinfo  {journal} {Physical Review Letters}\ }\textbf {\bibinfo
  {volume} {106}},\ \bibinfo {pages} {015302} (\bibinfo {year}
  {2011})}\BibitemShut {NoStop}%
\bibitem [{\citenamefont {\"{O}lschl\"{a}ger}\ \emph
  {et~al.}(2012)\citenamefont {\"{O}lschl\"{a}ger}, \citenamefont {Wirth},
  \citenamefont {Kock},\ and\ \citenamefont
  {Hemmerich}}]{olschlager_topologically_2012}%
  \BibitemOpen
  \bibfield  {author} {\bibinfo {author} {\bibfnamefont {M.}~\bibnamefont
  {\"{O}lschl\"{a}ger}}, \bibinfo {author} {\bibfnamefont {G.}~\bibnamefont
  {Wirth}}, \bibinfo {author} {\bibfnamefont {T.}~\bibnamefont {Kock}}, \ and\
  \bibinfo {author} {\bibfnamefont {A.}~\bibnamefont {Hemmerich}},\ }\href
  {\doibase 10.1103/PhysRevLett.108.075302} {\bibfield  {journal} {\bibinfo
  {journal} {Physical Review Letters}\ }\textbf {\bibinfo {volume} {108}},\
  \bibinfo {pages} {075302} (\bibinfo {year} {2012})}\BibitemShut {NoStop}%
\bibitem [{\citenamefont {\"{O}lschl\"{a}ger}\ \emph
  {et~al.}(2013)\citenamefont {\"{O}lschl\"{a}ger}, \citenamefont {Kock},
  \citenamefont {Wirth}, \citenamefont {Ewerbeck}, \citenamefont {Smith},\ and\
  \citenamefont {Hemmerich}}]{olschlager_interaction-induced_2013}%
  \BibitemOpen
  \bibfield  {author} {\bibinfo {author} {\bibfnamefont {M.}~\bibnamefont
  {\"{O}lschl\"{a}ger}}, \bibinfo {author} {\bibfnamefont {T.}~\bibnamefont
  {Kock}}, \bibinfo {author} {\bibfnamefont {G.}~\bibnamefont {Wirth}},
  \bibinfo {author} {\bibfnamefont {A.}~\bibnamefont {Ewerbeck}}, \bibinfo
  {author} {\bibfnamefont {C.~M.}\ \bibnamefont {Smith}}, \ and\ \bibinfo
  {author} {\bibfnamefont {A.}~\bibnamefont {Hemmerich}},\ }\href {\doibase
  10.1088/1367-2630/15/8/083041} {\bibfield  {journal} {\bibinfo  {journal}
  {New Journal of Physics}\ }\textbf {\bibinfo {volume} {15}},\ \bibinfo
  {pages} {083041} (\bibinfo {year} {2013})}\BibitemShut {NoStop}%
\bibitem [{\citenamefont {Zhao}\ and\ \citenamefont
  {Liu}(2008)}]{zhao_orbital_2008}%
  \BibitemOpen
  \bibfield  {author} {\bibinfo {author} {\bibfnamefont {E.}~\bibnamefont
  {Zhao}}\ and\ \bibinfo {author} {\bibfnamefont {W.~V.}\ \bibnamefont {Liu}},\
  }\href {\doibase 10.1103/PhysRevLett.100.160403} {\bibfield  {journal}
  {\bibinfo  {journal} {Physical Review Letters}\ }\textbf {\bibinfo {volume}
  {100}},\ \bibinfo {pages} {160403} (\bibinfo {year} {2008})}\BibitemShut
  {NoStop}%
\bibitem [{\citenamefont {Wu}(2008)}]{wu_orbital_2008}%
  \BibitemOpen
  \bibfield  {author} {\bibinfo {author} {\bibfnamefont {C.}~\bibnamefont
  {Wu}},\ }\href {\doibase 10.1103/PhysRevLett.100.200406} {\bibfield
  {journal} {\bibinfo  {journal} {Physical Review Letters}\ }\textbf {\bibinfo
  {volume} {100}},\ \bibinfo {pages} {200406} (\bibinfo {year}
  {2008})}\BibitemShut {NoStop}%
\bibitem [{\citenamefont {Wu}\ and\ \citenamefont
  {Zhai}(2008)}]{wu_theory_2008}%
  \BibitemOpen
  \bibfield  {author} {\bibinfo {author} {\bibfnamefont {K.}~\bibnamefont
  {Wu}}\ and\ \bibinfo {author} {\bibfnamefont {H.}~\bibnamefont {Zhai}},\
  }\href {\doibase 10.1103/PhysRevB.77.174431} {\bibfield  {journal} {\bibinfo
  {journal} {Physical Review B}\ }\textbf {\bibinfo {volume} {77}},\ \bibinfo
  {pages} {174431} (\bibinfo {year} {2008})}\BibitemShut {NoStop}%
\bibitem [{\citenamefont {Wang}\ \emph {et~al.}(2008)\citenamefont {Wang},
  \citenamefont {Dai}, \citenamefont {Chen},\ and\ \citenamefont
  {Xie}}]{wang_magnetism_2008}%
  \BibitemOpen
  \bibfield  {author} {\bibinfo {author} {\bibfnamefont {L.}~\bibnamefont
  {Wang}}, \bibinfo {author} {\bibfnamefont {X.}~\bibnamefont {Dai}}, \bibinfo
  {author} {\bibfnamefont {S.}~\bibnamefont {Chen}}, \ and\ \bibinfo {author}
  {\bibfnamefont {X.~C.}\ \bibnamefont {Xie}},\ }\href {\doibase
  10.1103/PhysRevA.78.023603} {\bibfield  {journal} {\bibinfo  {journal}
  {Physical Review A}\ }\textbf {\bibinfo {volume} {78}},\ \bibinfo {pages}
  {023603} (\bibinfo {year} {2008})}\BibitemShut {NoStop}%
\bibitem [{\citenamefont {Zhang}\ \emph {et~al.}(2010)\citenamefont {Zhang},
  \citenamefont {Hung},\ and\ \citenamefont {Wu}}]{zhang_proposed_2010}%
  \BibitemOpen
  \bibfield  {author} {\bibinfo {author} {\bibfnamefont {S.}~\bibnamefont
  {Zhang}}, \bibinfo {author} {\bibfnamefont {H.-h.}\ \bibnamefont {Hung}}, \
  and\ \bibinfo {author} {\bibfnamefont {C.}~\bibnamefont {Wu}},\ }\href
  {\doibase 10.1103/PhysRevA.82.053618} {\bibfield  {journal} {\bibinfo
  {journal} {Physical Review A}\ }\textbf {\bibinfo {volume} {82}},\ \bibinfo
  {pages} {053618} (\bibinfo {year} {2010})}\BibitemShut {NoStop}%
\bibitem [{\citenamefont {Li}\ \emph {et~al.}(2014)\citenamefont {Li},
  \citenamefont {Lieb},\ and\ \citenamefont {Wu}}]{li_exact_2014}%
  \BibitemOpen
  \bibfield  {author} {\bibinfo {author} {\bibfnamefont {Y.}~\bibnamefont
  {Li}}, \bibinfo {author} {\bibfnamefont {E.~H.}\ \bibnamefont {Lieb}}, \ and\
  \bibinfo {author} {\bibfnamefont {C.}~\bibnamefont {Wu}},\ }\href {\doibase
  10.1103/PhysRevLett.112.217201} {\bibfield  {journal} {\bibinfo  {journal}
  {Physical Review Letters}\ }\textbf {\bibinfo {volume} {112}},\ \bibinfo
  {pages} {217201} (\bibinfo {year} {2014})}\BibitemShut {NoStop}%
\bibitem [{\citenamefont {Wu}()}]{priv}%
  \BibitemOpen
  \bibfield  {author} {\bibinfo {author} {\bibfnamefont {C.}~\bibnamefont
  {Wu}},\ }\href@noop {} {}\bibinfo {note} {Private communication
  (unpublished)}\BibitemShut {NoStop}%
\bibitem [{\citenamefont {Fedichev}\ \emph {et~al.}(1996)\citenamefont
  {Fedichev}, \citenamefont {Kagan}, \citenamefont {Shlyapnikov},\ and\
  \citenamefont {Walraven}}]{PhysRevLett.77.2913}%
  \BibitemOpen
  \bibfield  {author} {\bibinfo {author} {\bibfnamefont {P.~O.}\ \bibnamefont
  {Fedichev}}, \bibinfo {author} {\bibfnamefont {Y.}~\bibnamefont {Kagan}},
  \bibinfo {author} {\bibfnamefont {G.~V.}\ \bibnamefont {Shlyapnikov}}, \ and\
  \bibinfo {author} {\bibfnamefont {J.~T.~M.}\ \bibnamefont {Walraven}},\
  }\href {\doibase 10.1103/PhysRevLett.77.2913} {\bibfield  {journal} {\bibinfo
   {journal} {Phys. Rev. Lett.}\ }\textbf {\bibinfo {volume} {77}},\ \bibinfo
  {pages} {2913} (\bibinfo {year} {1996})}\BibitemShut {NoStop}%
\bibitem [{\citenamefont {Bohn}\ and\ \citenamefont
  {Julienne}(1997)}]{PhysRevA.56.1486}%
  \BibitemOpen
  \bibfield  {author} {\bibinfo {author} {\bibfnamefont {J.~L.}\ \bibnamefont
  {Bohn}}\ and\ \bibinfo {author} {\bibfnamefont {P.~S.}\ \bibnamefont
  {Julienne}},\ }\href {\doibase 10.1103/PhysRevA.56.1486} {\bibfield
  {journal} {\bibinfo  {journal} {Phys. Rev. A}\ }\textbf {\bibinfo {volume}
  {56}},\ \bibinfo {pages} {1486} (\bibinfo {year} {1997})}\BibitemShut
  {NoStop}%
\bibitem [{\citenamefont {Fatemi}\ \emph {et~al.}(2000)\citenamefont {Fatemi},
  \citenamefont {Jones},\ and\ \citenamefont {Lett}}]{PhysRevLett.85.4462}%
  \BibitemOpen
  \bibfield  {author} {\bibinfo {author} {\bibfnamefont {F.~K.}\ \bibnamefont
  {Fatemi}}, \bibinfo {author} {\bibfnamefont {K.~M.}\ \bibnamefont {Jones}}, \
  and\ \bibinfo {author} {\bibfnamefont {P.~D.}\ \bibnamefont {Lett}},\ }\href
  {\doibase 10.1103/PhysRevLett.85.4462} {\bibfield  {journal} {\bibinfo
  {journal} {Phys. Rev. Lett.}\ }\textbf {\bibinfo {volume} {85}},\ \bibinfo
  {pages} {4462} (\bibinfo {year} {2000})}\BibitemShut {NoStop}%
\bibitem [{\citenamefont {Theis}\ \emph {et~al.}(2004)\citenamefont {Theis},
  \citenamefont {Thalhammer}, \citenamefont {Winkler}, \citenamefont {Hellwig},
  \citenamefont {Ruff}, \citenamefont {Grimm},\ and\ \citenamefont
  {Denschlag}}]{PhysRevLett.93.123001}%
  \BibitemOpen
  \bibfield  {author} {\bibinfo {author} {\bibfnamefont {M.}~\bibnamefont
  {Theis}}, \bibinfo {author} {\bibfnamefont {G.}~\bibnamefont {Thalhammer}},
  \bibinfo {author} {\bibfnamefont {K.}~\bibnamefont {Winkler}}, \bibinfo
  {author} {\bibfnamefont {M.}~\bibnamefont {Hellwig}}, \bibinfo {author}
  {\bibfnamefont {G.}~\bibnamefont {Ruff}}, \bibinfo {author} {\bibfnamefont
  {R.}~\bibnamefont {Grimm}}, \ and\ \bibinfo {author} {\bibfnamefont {J.~H.}\
  \bibnamefont {Denschlag}},\ }\href {\doibase 10.1103/PhysRevLett.93.123001}
  {\bibfield  {journal} {\bibinfo  {journal} {Phys. Rev. Lett.}\ }\textbf
  {\bibinfo {volume} {93}},\ \bibinfo {pages} {123001} (\bibinfo {year}
  {2004})}\BibitemShut {NoStop}%
\bibitem [{\citenamefont {Yamazaki}\ \emph {et~al.}(2010)\citenamefont
  {Yamazaki}, \citenamefont {Taie}, \citenamefont {Sugawa},\ and\ \citenamefont
  {Takahashi}}]{PhysRevLett.105.050405}%
  \BibitemOpen
  \bibfield  {author} {\bibinfo {author} {\bibfnamefont {R.}~\bibnamefont
  {Yamazaki}}, \bibinfo {author} {\bibfnamefont {S.}~\bibnamefont {Taie}},
  \bibinfo {author} {\bibfnamefont {S.}~\bibnamefont {Sugawa}}, \ and\ \bibinfo
  {author} {\bibfnamefont {Y.}~\bibnamefont {Takahashi}},\ }\href {\doibase
  10.1103/PhysRevLett.105.050405} {\bibfield  {journal} {\bibinfo  {journal}
  {Phys. Rev. Lett.}\ }\textbf {\bibinfo {volume} {105}},\ \bibinfo {pages}
  {050405} (\bibinfo {year} {2010})}\BibitemShut {NoStop}%
\bibitem [{\citenamefont {Blatt}\ \emph {et~al.}(2011)\citenamefont {Blatt},
  \citenamefont {Nicholson}, \citenamefont {Bloom}, \citenamefont {Williams},
  \citenamefont {Thomsen}, \citenamefont {Julienne},\ and\ \citenamefont
  {Ye}}]{PhysRevLett.107.073202}%
  \BibitemOpen
  \bibfield  {author} {\bibinfo {author} {\bibfnamefont {S.}~\bibnamefont
  {Blatt}}, \bibinfo {author} {\bibfnamefont {T.~L.}\ \bibnamefont
  {Nicholson}}, \bibinfo {author} {\bibfnamefont {B.~J.}\ \bibnamefont
  {Bloom}}, \bibinfo {author} {\bibfnamefont {J.~R.}\ \bibnamefont {Williams}},
  \bibinfo {author} {\bibfnamefont {J.~W.}\ \bibnamefont {Thomsen}}, \bibinfo
  {author} {\bibfnamefont {P.~S.}\ \bibnamefont {Julienne}}, \ and\ \bibinfo
  {author} {\bibfnamefont {J.}~\bibnamefont {Ye}},\ }\href {\doibase
  10.1103/PhysRevLett.107.073202} {\bibfield  {journal} {\bibinfo  {journal}
  {Phys. Rev. Lett.}\ }\textbf {\bibinfo {volume} {107}},\ \bibinfo {pages}
  {073202} (\bibinfo {year} {2011})}\BibitemShut {NoStop}%
\bibitem [{\citenamefont {Deb}\ and\ \citenamefont
  {Hazra}(2009)}]{PhysRevLett.103.023201}%
  \BibitemOpen
  \bibfield  {author} {\bibinfo {author} {\bibfnamefont {B.}~\bibnamefont
  {Deb}}\ and\ \bibinfo {author} {\bibfnamefont {J.}~\bibnamefont {Hazra}},\
  }\href {\doibase 10.1103/PhysRevLett.103.023201} {\bibfield  {journal}
  {\bibinfo  {journal} {Phys. Rev. Lett.}\ }\textbf {\bibinfo {volume} {103}},\
  \bibinfo {pages} {023201} (\bibinfo {year} {2009})}\BibitemShut {NoStop}%
\bibitem [{\citenamefont {Goyal}\ \emph {et~al.}(2010)\citenamefont {Goyal},
  \citenamefont {Reichenbach},\ and\ \citenamefont
  {Deutsch}}]{PhysRevA.82.062704}%
  \BibitemOpen
  \bibfield  {author} {\bibinfo {author} {\bibfnamefont {K.}~\bibnamefont
  {Goyal}}, \bibinfo {author} {\bibfnamefont {I.}~\bibnamefont {Reichenbach}},
  \ and\ \bibinfo {author} {\bibfnamefont {I.}~\bibnamefont {Deutsch}},\ }\href
  {\doibase 10.1103/PhysRevA.82.062704} {\bibfield  {journal} {\bibinfo
  {journal} {Phys. Rev. A}\ }\textbf {\bibinfo {volume} {82}},\ \bibinfo
  {pages} {062704} (\bibinfo {year} {2010})}\BibitemShut {NoStop}%
\bibitem [{\citenamefont {Yamazaki}\ \emph {et~al.}(2013)\citenamefont
  {Yamazaki}, \citenamefont {Taie}, \citenamefont {Sugawa}, \citenamefont
  {Enomoto},\ and\ \citenamefont {Takahashi}}]{PhysRevA.87.010704}%
  \BibitemOpen
  \bibfield  {author} {\bibinfo {author} {\bibfnamefont {R.}~\bibnamefont
  {Yamazaki}}, \bibinfo {author} {\bibfnamefont {S.}~\bibnamefont {Taie}},
  \bibinfo {author} {\bibfnamefont {S.}~\bibnamefont {Sugawa}}, \bibinfo
  {author} {\bibfnamefont {K.}~\bibnamefont {Enomoto}}, \ and\ \bibinfo
  {author} {\bibfnamefont {Y.}~\bibnamefont {Takahashi}},\ }\href {\doibase
  10.1103/PhysRevA.87.010704} {\bibfield  {journal} {\bibinfo  {journal} {Phys.
  Rev. A}\ }\textbf {\bibinfo {volume} {87}},\ \bibinfo {pages} {010704}
  (\bibinfo {year} {2013})}\BibitemShut {NoStop}%
\bibitem [{\citenamefont {Chern}\ and\ \citenamefont
  {Wu}(2011)}]{PhysRevE.84.061127}%
  \BibitemOpen
  \bibfield  {author} {\bibinfo {author} {\bibfnamefont {G.-W.}\ \bibnamefont
  {Chern}}\ and\ \bibinfo {author} {\bibfnamefont {C.}~\bibnamefont {Wu}},\
  }\href {\doibase 10.1103/PhysRevE.84.061127} {\bibfield  {journal} {\bibinfo
  {journal} {Phys. Rev. E}\ }\textbf {\bibinfo {volume} {84}},\ \bibinfo
  {pages} {061127} (\bibinfo {year} {2011})}\BibitemShut {NoStop}%
\bibitem [{\citenamefont {Hauke}\ \emph {et~al.}(2011)\citenamefont {Hauke},
  \citenamefont {Zhao}, \citenamefont {Goyal}, \citenamefont {Deutsch},
  \citenamefont {Liu},\ and\ \citenamefont {Lewenstein}}]{PhysRevA.84.051603}%
  \BibitemOpen
  \bibfield  {author} {\bibinfo {author} {\bibfnamefont {P.}~\bibnamefont
  {Hauke}}, \bibinfo {author} {\bibfnamefont {E.}~\bibnamefont {Zhao}},
  \bibinfo {author} {\bibfnamefont {K.}~\bibnamefont {Goyal}}, \bibinfo
  {author} {\bibfnamefont {I.~H.}\ \bibnamefont {Deutsch}}, \bibinfo {author}
  {\bibfnamefont {W.~V.}\ \bibnamefont {Liu}}, \ and\ \bibinfo {author}
  {\bibfnamefont {M.}~\bibnamefont {Lewenstein}},\ }\href {\doibase
  10.1103/PhysRevA.84.051603} {\bibfield  {journal} {\bibinfo  {journal} {Phys.
  Rev. A}\ }\textbf {\bibinfo {volume} {84}},\ \bibinfo {pages} {051603}
  (\bibinfo {year} {2011})}\BibitemShut {NoStop}%
\bibitem [{\citenamefont {Liu}\ and\ \citenamefont
  {Wu}(2006)}]{liu_atomic_2006}%
  \BibitemOpen
  \bibfield  {author} {\bibinfo {author} {\bibfnamefont {W.~V.}\ \bibnamefont
  {Liu}}\ and\ \bibinfo {author} {\bibfnamefont {C.}~\bibnamefont {Wu}},\
  }\href {\doibase 10.1103/PhysRevA.74.013607} {\bibfield  {journal} {\bibinfo
  {journal} {Physical Review A}\ }\textbf {\bibinfo {volume} {74}},\ \bibinfo
  {pages} {013607} (\bibinfo {year} {2006})}\BibitemShut {NoStop}%
\bibitem [{\citenamefont {Ole\'{s}}\ \emph {et~al.}(2000)\citenamefont
  {Ole\'{s}}, \citenamefont {Felix~Feiner},\ and\ \citenamefont
  {Zaanen}}]{oles_quantum_2000}%
  \BibitemOpen
  \bibfield  {author} {\bibinfo {author} {\bibfnamefont {A.~M.}\ \bibnamefont
  {Ole\'{s}}}, \bibinfo {author} {\bibfnamefont {L.}~\bibnamefont
  {Felix~Feiner}}, \ and\ \bibinfo {author} {\bibfnamefont {J.}~\bibnamefont
  {Zaanen}},\ }\href {\doibase 10.1103/PhysRevB.61.6257} {\bibfield  {journal}
  {\bibinfo  {journal} {Physical Review B}\ }\textbf {\bibinfo {volume} {61}},\
  \bibinfo {pages} {6257} (\bibinfo {year} {2000})}\BibitemShut {NoStop}%
\bibitem [{\citenamefont {Li}\ \emph {et~al.}(1998)\citenamefont {Li},
  \citenamefont {Ma}, \citenamefont {Shi},\ and\ \citenamefont
  {Zhang}}]{li_su4_1998}%
  \BibitemOpen
  \bibfield  {author} {\bibinfo {author} {\bibfnamefont {Y.~Q.}\ \bibnamefont
  {Li}}, \bibinfo {author} {\bibfnamefont {M.}~\bibnamefont {Ma}}, \bibinfo
  {author} {\bibfnamefont {D.~N.}\ \bibnamefont {Shi}}, \ and\ \bibinfo
  {author} {\bibfnamefont {F.~C.}\ \bibnamefont {Zhang}},\ }\href {\doibase
  10.1103/PhysRevLett.81.3527} {\bibfield  {journal} {\bibinfo  {journal}
  {Physical Review Letters}\ }\textbf {\bibinfo {volume} {81}},\ \bibinfo
  {pages} {3527} (\bibinfo {year} {1998})}\BibitemShut {NoStop}%
\bibitem [{\citenamefont {Yamashita}\ \emph {et~al.}(1998)\citenamefont
  {Yamashita}, \citenamefont {Shibata},\ and\ \citenamefont
  {Ueda}}]{yamashita_su4_1998}%
  \BibitemOpen
  \bibfield  {author} {\bibinfo {author} {\bibfnamefont {Y.}~\bibnamefont
  {Yamashita}}, \bibinfo {author} {\bibfnamefont {N.}~\bibnamefont {Shibata}},
  \ and\ \bibinfo {author} {\bibfnamefont {K.}~\bibnamefont {Ueda}},\ }\href
  {\doibase 10.1103/PhysRevB.58.9114} {\bibfield  {journal} {\bibinfo
  {journal} {Physical Review B}\ }\textbf {\bibinfo {volume} {58}},\ \bibinfo
  {pages} {9114} (\bibinfo {year} {1998})}\BibitemShut {NoStop}%
\bibitem [{\citenamefont {Wang}\ and\ \citenamefont
  {Vishwanath}(2009)}]{wang_z2_2009}%
  \BibitemOpen
  \bibfield  {author} {\bibinfo {author} {\bibfnamefont {F.}~\bibnamefont
  {Wang}}\ and\ \bibinfo {author} {\bibfnamefont {A.}~\bibnamefont
  {Vishwanath}},\ }\href {\doibase 10.1103/PhysRevB.80.064413} {\bibfield
  {journal} {\bibinfo  {journal} {Physical Review B}\ }\textbf {\bibinfo
  {volume} {80}},\ \bibinfo {pages} {064413} (\bibinfo {year}
  {2009})}\BibitemShut {NoStop}%
\bibitem [{\citenamefont {Kugel'}\ and\ \citenamefont
  {Khomski\u{i}}(1973)}]{kugel_crystal_1973}%
  \BibitemOpen
  \bibfield  {author} {\bibinfo {author} {\bibfnamefont {K.~I.}\ \bibnamefont
  {Kugel'}}\ and\ \bibinfo {author} {\bibfnamefont {D.~I.}\ \bibnamefont
  {Khomski\u{i}}},\ }\href {http://adsabs.harvard.edu/abs/1973JETP...37..725K}
  {\bibfield  {journal} {\bibinfo  {journal} {Soviet Journal of Experimental
  and Theoretical Physics}\ }\textbf {\bibinfo {volume} {37}},\ \bibinfo
  {pages} {725} (\bibinfo {year} {1973})}\BibitemShut {NoStop}%
\bibitem [{\citenamefont {Gorshkov}\ \emph {et~al.}(2010)\citenamefont
  {Gorshkov}, \citenamefont {Hermele}, \citenamefont {Gurarie}, \citenamefont
  {Xu}, \citenamefont {Julienne}, \citenamefont {Ye}, \citenamefont {Zoller},
  \citenamefont {Demler}, \citenamefont {Lukin},\ and\ \citenamefont
  {Rey}}]{gorshkov_two-orbital_2010}%
  \BibitemOpen
  \bibfield  {author} {\bibinfo {author} {\bibfnamefont {A.~V.}\ \bibnamefont
  {Gorshkov}}, \bibinfo {author} {\bibfnamefont {M.}~\bibnamefont {Hermele}},
  \bibinfo {author} {\bibfnamefont {V.}~\bibnamefont {Gurarie}}, \bibinfo
  {author} {\bibfnamefont {C.}~\bibnamefont {Xu}}, \bibinfo {author}
  {\bibfnamefont {P.~S.}\ \bibnamefont {Julienne}}, \bibinfo {author}
  {\bibfnamefont {J.}~\bibnamefont {Ye}}, \bibinfo {author} {\bibfnamefont
  {P.}~\bibnamefont {Zoller}}, \bibinfo {author} {\bibfnamefont
  {E.}~\bibnamefont {Demler}}, \bibinfo {author} {\bibfnamefont {M.~D.}\
  \bibnamefont {Lukin}}, \ and\ \bibinfo {author} {\bibfnamefont {A.~M.}\
  \bibnamefont {Rey}},\ }\href {\doibase 10.1038/nphys1535} {\bibfield
  {journal} {\bibinfo  {journal} {Nature Physics}\ }\textbf {\bibinfo {volume}
  {6}},\ \bibinfo {pages} {289} (\bibinfo {year} {2010})}\BibitemShut {NoStop}%
\bibitem [{\citenamefont {Bethe}(1931)}]{bethe_zur_1931}%
  \BibitemOpen
  \bibfield  {author} {\bibinfo {author} {\bibfnamefont {H.}~\bibnamefont
  {Bethe}},\ }\href {\doibase 10.1007/BF01341708} {\bibfield  {journal}
  {\bibinfo  {journal} {Zeitschrift f{\"u}r Physik}\ }\textbf {\bibinfo
  {volume} {71}},\ \bibinfo {pages} {205} (\bibinfo {year} {1931})}\BibitemShut
  {NoStop}%
\bibitem [{\citenamefont {Imambekov}\ \emph {et~al.}(2012)\citenamefont
  {Imambekov}, \citenamefont {Schmidt},\ and\ \citenamefont
  {Glazman}}]{RevModPhys.84.1253}%
  \BibitemOpen
  \bibfield  {author} {\bibinfo {author} {\bibfnamefont {A.}~\bibnamefont
  {Imambekov}}, \bibinfo {author} {\bibfnamefont {T.~L.}\ \bibnamefont
  {Schmidt}}, \ and\ \bibinfo {author} {\bibfnamefont {L.~I.}\ \bibnamefont
  {Glazman}},\ }\href {\doibase 10.1103/RevModPhys.84.1253} {\bibfield
  {journal} {\bibinfo  {journal} {Rev. Mod. Phys.}\ }\textbf {\bibinfo {volume}
  {84}},\ \bibinfo {pages} {1253} (\bibinfo {year} {2012})}\BibitemShut
  {NoStop}%
\bibitem [{\citenamefont {Kinoshita}\ \emph {et~al.}(2006)\citenamefont
  {Kinoshita}, \citenamefont {Wenger},\ and\ \citenamefont
  {Weiss}}]{kinoshita2006quantum}%
  \BibitemOpen
  \bibfield  {author} {\bibinfo {author} {\bibfnamefont {T.}~\bibnamefont
  {Kinoshita}}, \bibinfo {author} {\bibfnamefont {T.}~\bibnamefont {Wenger}}, \
  and\ \bibinfo {author} {\bibfnamefont {D.~S.}\ \bibnamefont {Weiss}},\
  }\href@noop {} {\bibfield  {journal} {\bibinfo  {journal} {Nature}\ }\textbf
  {\bibinfo {volume} {440}},\ \bibinfo {pages} {900} (\bibinfo {year}
  {2006})}\BibitemShut {NoStop}%
\bibitem [{\citenamefont {Rigol}\ \emph {et~al.}(2008)\citenamefont {Rigol},
  \citenamefont {Dunjko},\ and\ \citenamefont
  {Olshanii}}]{rigol2008thermalization}%
  \BibitemOpen
  \bibfield  {author} {\bibinfo {author} {\bibfnamefont {M.}~\bibnamefont
  {Rigol}}, \bibinfo {author} {\bibfnamefont {V.}~\bibnamefont {Dunjko}}, \
  and\ \bibinfo {author} {\bibfnamefont {M.}~\bibnamefont {Olshanii}},\
  }\href@noop {} {\bibfield  {journal} {\bibinfo  {journal} {Nature}\ }\textbf
  {\bibinfo {volume} {452}},\ \bibinfo {pages} {854} (\bibinfo {year}
  {2008})}\BibitemShut {NoStop}%
\bibitem [{\citenamefont {Chakravarty}(1996)}]{PhysRevLett.77.4446}%
  \BibitemOpen
  \bibfield  {author} {\bibinfo {author} {\bibfnamefont {S.}~\bibnamefont
  {Chakravarty}},\ }\href {\doibase 10.1103/PhysRevLett.77.4446} {\bibfield
  {journal} {\bibinfo  {journal} {Phys. Rev. Lett.}\ }\textbf {\bibinfo
  {volume} {77}},\ \bibinfo {pages} {4446} (\bibinfo {year}
  {1996})}\BibitemShut {NoStop}%
\bibitem [{\citenamefont {Affleck}\ and\ \citenamefont
  {Halperin}(1996)}]{0305-4470-29-11-003}%
  \BibitemOpen
  \bibfield  {author} {\bibinfo {author} {\bibfnamefont {I.}~\bibnamefont
  {Affleck}}\ and\ \bibinfo {author} {\bibfnamefont {B.~I.}\ \bibnamefont
  {Halperin}},\ }\href {http://stacks.iop.org/0305-4470/29/i=11/a=003}
  {\bibfield  {journal} {\bibinfo  {journal} {Journal of Physics A:
  Mathematical and General}\ }\textbf {\bibinfo {volume} {29}},\ \bibinfo
  {pages} {2627} (\bibinfo {year} {1996})}\BibitemShut {NoStop}%
\bibitem [{\citenamefont {Lee}\ \emph {et~al.}(2006)\citenamefont {Lee},
  \citenamefont {Park}, \citenamefont {Adroja}, \citenamefont {Khomskii},
  \citenamefont {Streltsov}, \citenamefont {McEwen}, \citenamefont {Sakai},
  \citenamefont {Yoshimura}, \citenamefont {Anisimov}, \citenamefont {Mori},
  \citenamefont {Kanno},\ and\ \citenamefont {Ibberson}}]{lee_spin_2006}%
  \BibitemOpen
  \bibfield  {author} {\bibinfo {author} {\bibfnamefont {S.}~\bibnamefont
  {Lee}}, \bibinfo {author} {\bibfnamefont {J.-G.}\ \bibnamefont {Park}},
  \bibinfo {author} {\bibfnamefont {D.~T.}\ \bibnamefont {Adroja}}, \bibinfo
  {author} {\bibfnamefont {D.}~\bibnamefont {Khomskii}}, \bibinfo {author}
  {\bibfnamefont {S.}~\bibnamefont {Streltsov}}, \bibinfo {author}
  {\bibfnamefont {K.~A.}\ \bibnamefont {McEwen}}, \bibinfo {author}
  {\bibfnamefont {H.}~\bibnamefont {Sakai}}, \bibinfo {author} {\bibfnamefont
  {K.}~\bibnamefont {Yoshimura}}, \bibinfo {author} {\bibfnamefont {V.~I.}\
  \bibnamefont {Anisimov}}, \bibinfo {author} {\bibfnamefont {D.}~\bibnamefont
  {Mori}}, \bibinfo {author} {\bibfnamefont {R.}~\bibnamefont {Kanno}}, \ and\
  \bibinfo {author} {\bibfnamefont {R.}~\bibnamefont {Ibberson}},\ }\href
  {\doibase 10.1038/nmat1605} {\bibfield  {journal} {\bibinfo  {journal}
  {Nature Materials}\ }\textbf {\bibinfo {volume} {5}},\ \bibinfo {pages} {471}
  (\bibinfo {year} {2006})}\BibitemShut {NoStop}%
\bibitem [{\citenamefont {van~den
  Brink}(2006)}]{van_den_brink_transition_2006}%
  \BibitemOpen
  \bibfield  {author} {\bibinfo {author} {\bibfnamefont {J.}~\bibnamefont
  {van~den Brink}},\ }\href {\doibase 10.1038/nmat1648} {\bibfield  {journal}
  {\bibinfo  {journal} {Nature Materials}\ }\textbf {\bibinfo {volume} {5}},\
  \bibinfo {pages} {427} (\bibinfo {year} {2006})}\BibitemShut {NoStop}%
\end{thebibliography}%

\end{document}